# Secure, Efficient Data Transport and Replica Management for High-Performance Data-Intensive Computing

Bill Allcock[1]  Joe Bester[1]  John Bresnahan[1]  Ann L. Chervenak[2]  Ian Foster[1,3]
Carl Kesselman[2]  Sam Meder[1]  Veronika Nefedova[1]  Darcy Quesnel[1]  Steven Tuecke[1]

[1] Mathematics and Computer Science Division
Argonne National Laboratory
Argonne, IL 60439
{allcock, bester, foster, nefedova, quesnel, tuecke}@mcs.anl.gov

[2] Information Sciences Institute
University of Southern California
Los Angeles, CA 90292
{annc, carl}@isi.edu

[3] Department of Computer Science & The Computation Institute
The University of Chicago
Chicago, IL 60637

**Abstract**

An emerging class of data-intensive applications involve the geographically dispersed extraction of complex scientific information from very large collections of measured or computed data. Such applications arise, for example, in experimental physics, where the data in question is generated by accelerators, and in simulation science, where the data is generated by supercomputers. So-called Data Grids provide essential infrastructure for such applications, much as the Internet provides essential services for applications such as e-mail and the Web. We describe here two services that we believe are fundamental to any Data Grid: reliable, high-speed transport and replica management. Our high-speed transport service, GridFTP, extends the popular FTP protocol with new features required for Data Grid applications, such as striping and partial file access. Our replica management service integrates a replica catalog with GridFTP transfers to provide for the creation, registration, location, and management of dataset replicas. We present the design of both services and also preliminary performance results. Our implementations exploit security and other services provided by the Globus Toolkit.

## 1 Introduction

Data-intensive, high-performance computing applications require the efficient management and transfer of terabytes or petabytes of information in wide-area, distributed computing environments. Examples of such applications include experimental analyses and simulations in scientific disciplines such as high-energy physics, climate modeling, earthquake engineering, and astronomy. In such applications, massive datasets must be shared by a community of hundreds or thousands of researchers distributed worldwide. These researchers need to be able to transfer large subsets of these datasets to local sites or other remote resources for processing. They may create local copies or replicas to overcome long wide-area data transfer latencies. The data management environment must provide security services such as authentication of users and control over who is allowed to access the data. In addition, once multiple copies of files are distributed at multiple locations, researchers need to be able to locate copies and determine whether to access an existing copy or create a new one to meet the performance needs of their applications.

We have argued elsewhere [1] that the requirements of such distributed data intensive applications are best met by the creation of a Data Grid infrastructure that provides a set of orthogonal, application-independent services that can then be combined and specialized in different ways to meet the needs of specific applications. We have argued further that this Data Grid infrastructure can usefully build on capabilities provided by the emerging Grid [2], such as resource access, resource discovery, and authentication services. Our Globus Toolkit [3] provides a widely used instantiation of the lower layers of this Grid architecture.

In this paper, we focus our attention on what we view as two fundamental Data Grid services, namely, *secure, reliable, efficient data transfer* and the ability to *register, locate, and manage multiple copies* of datasets. We describe the design, prototype implementation, and preliminary performance evaluation of our realization of these two services within the context of the Globus Toolkit. Given these two services, a wide range of higher-level data management services can be constructed, including reliable creation of a copy of a large data collection at a new location; selection of the best replica for a data transfer based on performance estimates provide by information services; and automatic creation of new replicas in response to application demands. However, we do not directly address these issues here.

## 2 Data-Intensive Computing Requirements

We use two application examples to motivate the design of our Data Grid services: high-energy physics experiments and climate modeling. We characterize each with respect to parameters such as average file sizes, total data volume, rate of data creation, types of file access (write-once, write-many), expected access rates, type of storage system (file system or database), and consistency requirements for multiple copies of data. In both these applications, as well as others that we have examined, such as earthquake engineering and astronomy, we see a common requirement for two basic data management services: efficient access to, and transfer of, large files; and a mechanism for creating and managing multiple copies of files.

### 2.1 High-Energy Physics Applications

Experimental physics applications operate on and generate large amounts of data. For example, beginning in 2005, the Large Hadron Collider (LHC) at the European physics center CERN will produce several petabytes of raw and derived data per year for approximately 15 years. The data generated by physics experiments is of two types: *experimental data*, or information collected *by* the experiment; and *metadata*, or information *about* the experiment, such as the number of events, and the results of analysis.

File sizes and numbers of files are determined to some extent by the type of software used to store experimental data and metadata. For example, several experiments have chosen to use the object-oriented Objectivity database. Current file sizes (e.g., within the BaBar experiment) range from 2 to 10 gigabytes in size, while metadata files are approximately 2 gigabytes. Objectivity currently limits database federations to 64K files. However, future versions of Objectivity will support more files, allowing average file sizes to be reduced.

Access patterns vary for experimental data files and metadata. Experimental data files typically have a single creator. During an initial production period lasting several weeks, these files are modified as new objects are added. After data production is complete, files are not modified. In contrast, metadata files may be created by multiple individuals and may be modified or augmented over time, even after the initial period of data production. For example, some experiments continue to modify metadata files to reflect the increasing number of total events in the database. The volume of metadata is typically smaller than that of experimental data.

The consumers of experimental physics data and metadata will number in the hundreds or thousands. These users are distributed at many sites worldwide. Hence, it is often desirable to make copies or *replicas* of the data being analyzed to minimize access time and network load. For example, Figure 1 shows the expected replication scheme for LHC physics datasets. Files are replicated in a hierarchical manner, with all files stored at a central location (CERN) and decreasing subsets of the data stored at national and regional data centers [4][5].

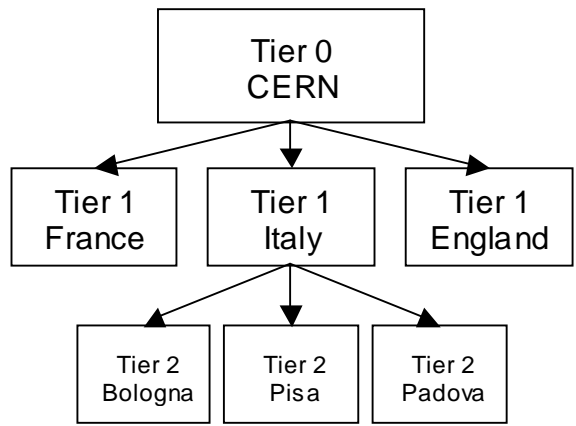

Figure 1: Scheme for hierarchical replication of Physics data

Replication of physics datasets is complicated by several factors. First, security services are required to authenticate the user and control access to storage systems. Next, because datasets are so large, it may be desirable to replicate only "interesting" subsets of the data. Finally, replication of data subject to modification implies a need for a mechanism for propagating updates to all replicas. For example, consider the initial period of data production, during which files are modified for several weeks. During this period, users want their local replicas to be updated periodically to reflect the experimental data being produced. Typically, updates are batched and performed every few days. Since metadata updates take place over an indefinite period, these changes must also be propagated periodically to all replicas.

In Table 1, we summarize the characteristics of high-energy physics applications.

Table 1: Characteristics of high-energy physics applications

| Rate of data generation (starting 2005) | Several petabytes per year |
|---|---|
| Typical experimental database file sizes | 2 to 10 gigabytes |
| Typical metadata database file sizes | 2 gigabytes |
| Maximum number of database files in federation | Currently 64K; eventually millions |
| Period of updates to experimental data | Several weeks |
| Period of updates to metadata | Indefinite |
| Type of storage system | Object-oriented database |
| Number of data consumers | Hundreds to thousands |

**2.2 Climate Modeling Application**

Climate modeling research groups generate large (multi-terabyte) *reference simulations* at supercomputer centers. These data are typically released in stages to progressively larger communities: first the research collaboration that generated the data, then perhaps selected colleagues, and eventually the entire community. To determine which users are allowed to view the collection at each stage, these applications require access control.

Reference simulation data are typically stored in a file system, often using a structured data format such as NetCDF, with associated metadata. Files are not updated once released. However, as in the physics application, climate modeling researchers find it convenient to create local copies of portions of the data. Therefore, the application has similar needs for managing copies of datasets at multiple locations, as well as for higher-level services such as replica selection or automatic replica creation.

**3 The Globus Toolkit**

The term *Grid computing* refers to the emerging computational and networking infrastructure that is designed to provide pervasive, uniform and reliable access to data, computational, and human resources distributed over wide area environments [6]. Grid services allow scientists at locations throughout the world to share data collection instruments such as particle colliders, compute resources such as supercomputers and clusters of workstations, and community datasets stored on network caches and hierarchical storage systems.

The Globus Toolkit developed within the Globus project provides middleware services for Grid computing environments. Major components include the Grid Security Infrastructure (GSI), which provides public-key-based authentication and authorization services; resource management services, which provide a language for specifying application requirements, mechanisms for immediate and advance reservations of Grid resources, and for remote job management; and information services, which provide for the distributed publication and retrieval of information about Grid resources.

Data Grid services complement and build on these components. For example, the GridFTP transfer service and the replica management service described in the rest of this paper use GSI for authentication and authorization. Higher-level data replication services can use the information service to locate the "best" replica and the resource management service to reserve the computational, network, and storage resources required by a data movement operation.

## 4  GridFTP:  A Secure, Efficient Data Transport Mechanism

The applications that we consider use a variety of storage systems, each designed to satisfy specific needs and requirements for storing, transferring and accessing large datasets. These include the Distributed Parallel Storage System (DPSS) and the High Performance Storage System (HPSS), which provide high-performance access to data and utilize parallel data transfer and/or striping across multiple servers to improve performance [7][8], and the Storage Resource Broker (SRB), which connects heterogeneous data collections, provides a uniform client interface to storage repositories, and provides a metadata catalog for describing and locating data within the storage system [9].

Unfortunately, these storage systems typically use incompatible and often unpublished protocols for accessing data, and therefore each requires the use of its own client. These incompatible protocols and client libraries effectively partition the datasets available on the Grid. Applications that require access to data stored in different storage systems must use multiple access methods.

To overcome these incompatible protocols, we propose a universal Grid data transfer and access protocol called *GridFTP* that provides secure, efficient data movement in Grid environments. This protocol, which extends the standard FTP protocol, provides a superset of the features offered by the various Grid storage systems currently in use. We argue that using GridFTP as a common data access protocol would be mutually advantageous to Grid storage providers and users. Storage providers gain a broader user base, because their data are available to any client, while storage users gain access to a broader range of storage systems and data.

We chose to extend the FTP protocol (rather than, for example, WebDAV) because we observed that FTP is the protocol most commonly used for data transfer on the Internet and the most likely candidate for meeting the Grid's needs. FTP is a widely implemented and well-understood IETF standard protocol with a large base of code and expertise from which to build. In addition, the FTP protocol provides a well-defined architecture for protocol extensions and supports dynamic discovery of the extensions supported by a particular implementation. Third, numerous groups have added extensions through the IETF, and some of these extensions are particularly useful in the Grid.

### 4.1 GridFTP Functionality

GridFTP functionality includes some features that are supported by FTP extensions that have already been standardized (RFC 959) but are seldom implemented in current systems. Other features are new extensions to FTP.

- **Grid Security Infrastructure and Kerberos support:** Robust and flexible authentication, integrity, and confidentiality features are critical when transferring or accessing files. GridFTP must support GSI and Kerberos authentication, with user controlled setting of various levels of data integrity and/or confidentiality. GridFTP immplements the authentication mechanisms defined by RFC 2228, "FTP Security Extensions".

- **Third-party control of data transfer:** To manage large datasets for distributed communities, we must provide authenticated *third-party* control of data transfers between storage servers. A third-party operation allows a user or application at one site to initiate, monitor and control a data transfer operation between two other sites: the source and destination for the data transfer. Our implementation adds Generic Security Services (GSS)-API authentication to the existing third-party transfer capability defined in the FTP standard.

- **Parallel data transfer:** On wide-area links, using multiple TCP streams in parallel (even between the same source and destination) can improve aggregate bandwidth over using a single TCP stream [10]. GridFTP supports parallel data transfer through FTP command extensions and data channel extensions.

- **Striped data transfer:** Data may be striped or interleaved across multiple servers, as in a DPSS network disk cache [11]. GridFTP includes extensions that initiate striped transfers, which use multiple TCP streams to transfer data that is partitioned among multiple servers. Striped transfers provide further bandwidth improvements over those achieved with parallel transfers. We have defined GridFTP protocol extensions that support striped data transfers.

- **Partial file transfer:** Some applications can benefit from transferring portions of files rather than complete files: for example, high-energy physics analyses that require access to relatively small subsets of massive, object-oriented physics database files. The best that the standard FTP protocol allows is transfer of the remainder of a file starting at a particular offset. GridFTP provides commands to support transfers of arbitrary subsets or regions of a file.

- **Automatic negotiation of TCP buffer/window sizes:** Using optimal settings for TCP buffer/window sizes can dramatically improve data transfer performance. However, manually setting TCP buffer/window sizes is an error-prone process (particularly for non-experts) and is often simply not done. GridFTP extends the standard FTP command set and data channel protocol to support both manual setting and automatic negotiation of TCP buffer sizes for large files and for large sets of small files.

- **Support for reliable and restartable data transfer:** Reliable transfer is important for many applications that manage data. Fault recovery methods are needed to handle failures such as transient network and server outages. The FTP standard includes basic features for restarting failed transfers that are not widely implemented. GridFTP exploits these features and extends them to cover the new data channel protocol.

### 4.2 The GridFTP Protocol Implementation

Our implementation of the GridFTP protocol supports partial file transfers, third-party transfers, parallel transfers and striped transfers. We do not yet support automatic negotiation of TCP buffer/window sizes. The implementation consists of two principal C libraries: the globus_ftp_control_library and the globus_ftp_client_library.

The **globus_ftp_control_library** implements the control channel API. This API provides routines for managing a GridFTP connection, including authentication, creation of control and data channels, and reading and writing data over data channels. Having separate control and data channels, as defined in the FTP protocol standard, greatly facilitates the support of such features as parallel, striped and third-party data transfers. For parallel and striped transfers, the control channel is used to specify a put or get operation; concurrent data transfer occurs over multiple parallel TCP data channels. In a third-party transfer, the initiator monitors or aborts the operation via the control channel, while data transfer is performed over one or more data channels between source and destination sites.

The **globus_ftp_client_library** implements the GridFTP client API. This API provides higher-level client features on top of the globus_ftp_control library, including complete file get and put operations, calls to set the level of parallelism for parallel data transfers, partial file transfer operations, third-party transfers, and eventually, functions to set TCP buffer sizes.

### 4.3 GridFTP Performance

Preliminary performance measurements of our GridFTP prototype demonstrate that we can indeed obtain high performance and reliable transfers in wide area networks. Further improvements are expected as a result of tuning and improvements to the implementation.

Figure 2 shows the performance of GridFTP transfers between two workstations, one at Argonne National Laboratory in Illinois and the other at Lawrence Berkeley National Laboratory in California, connected over the ES-Net network (www.es.net). Both workstations run the Linux operating system and have RAID storage systems with read/write bandwidth of approximately 60 megabytes per second. Gigabit Ethernet is the slowest link in the network path. The bottom curve in the graph shows GridFTP performance as the number of simultaneous TCP streams increases. For comparison, the top curve in the graph shows the performance of the same number of TCP streams measured by iperf, a tool for evaluating network performance that performs no disk I/O and has minimal CPU or protocol overhead [12]. Iperf provides one measurement of the maximum throughput of the network. Our experiment was run in random order relative to the number of streams, with the GridFTP measurement for a certain number of streams followed immediately by the iperf measurement for the same number of streams. For example, we took GridFTP measurements followed by iperf measurements for 18 simultaneous streams, then we took the two measurements for five simultaneous streams, etc. This randomization prevents any system or network trends from biasing the results, but assures that iperf and GridFTP measurements for the same number of streams are run close together temporally to reflect possible interactions with the number of streams. Each data point on the graph represents a single measurement. Iperf measurements were made using a one megabyte window size and ran for 30 seconds. GridFTP measurements recorded the time to transfer a one gigabyte file.

The graph shows that GridFTP bandwidth increases with the number of parallel TCP streams between the two workstations, until bandwidth reaches about 200 megabits per second with seven to ten TCP streams. Differences between iperf and GridFTP performance can be attributed to overheads including authentication and protocol operations, reporting performance to the client, and checkpointing data to allow restart. GridFTP achieves on average approximately 78% of the iperf bandwidth, although there is a great deal of variability. We speculate that this variability is due to the requirement that GridFTP wait for any packets that are misrouted or dropped, while iperf simply runs for its allotted time and stops regardless of whether there are outstanding packets.

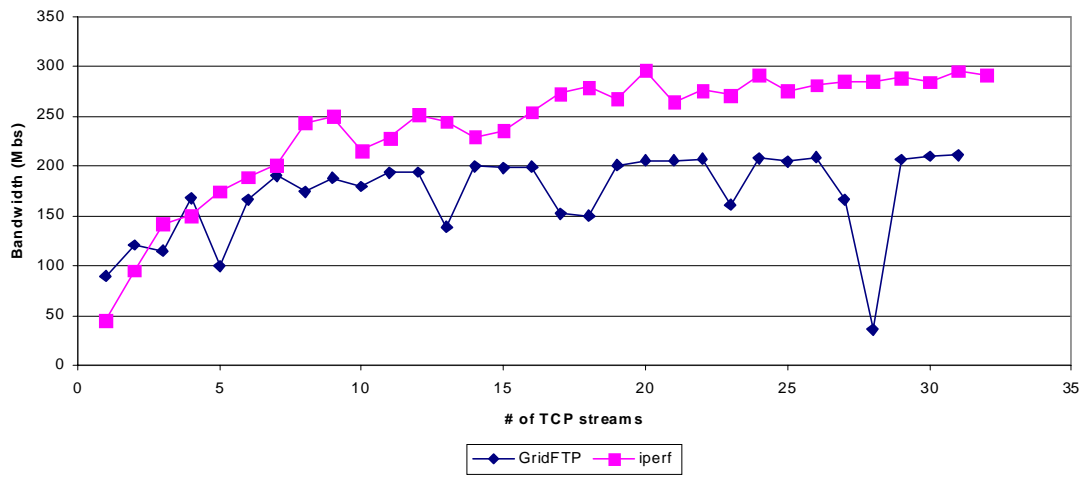

Figure 2: GridFTP performance compared to iperf measurements of network connection between Argonne National Laboratory and Lawrence Berkeley National Laboratory.

Figure 3 demonstrates GridFTP reliability. We show aggregate parallel bandwidth for a period of approximately fourteen hours during the SC'00 Conference in Dallas, Texas, on November 7, 2000. This data corresponds to parallel transfers between two uniprocessor hosts using varying levels of parallelism, up to a maximum of eight streams. The graph was produced with the NetLogger system [13]. Bandwidth between the two hosts reaches approximately 80 megabits per second, somewhat lower than shown for the hosts in Figure 2, most likely due to disk bandwidth limitations. Figure 3 shows drops in performance due to various network problems, including a power failure for the SC network (SCiNet), DNS problems, and backbone problems on the exhibition floor. Because the GridFTP protocol supports restart of failed transfers, the interrupted transfers are able to continue as soon as the network is restored. Toward the right side of the graph, we see several temporary increases in aggregate bandwidth, due to increased levels of parallelism. The frequent drop in bandwidth to relatively low levels occurs because our current implementation of GridFTP destroys and rebuilds its TCP connections between consecutive transfers. To address this problem, our next GridFTP implementation will support *data channel caching*. This mechanism allows a client to indicate that a TCP stream is likely to be re-used soon after the existing transfer is complete. In response to this hint, we will temporarily keep the TCP channel active and allow subsequent transfers to use the channel without requiring costly breakdown, restart, and re-authentication operations.

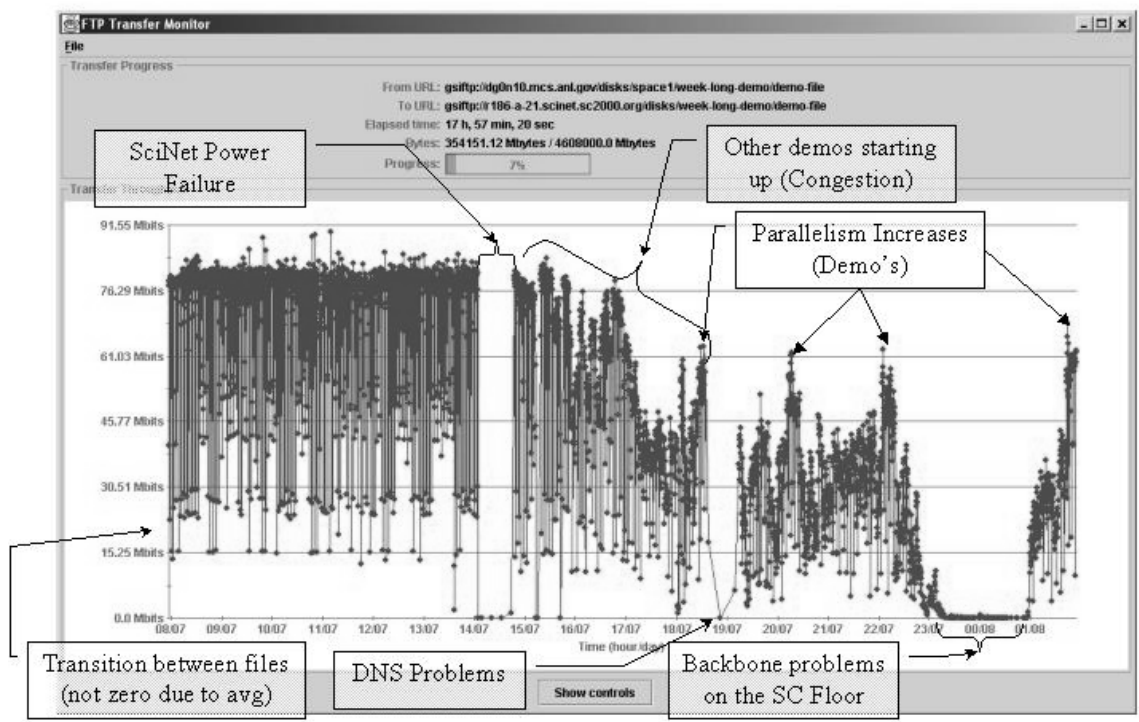

Figure 3: Bandwidth measured for a series of transfers performed over a 14 hour period, between Dallas and Chicago

Figure 4 and Table 2 address our achievable peak performance. These data were obtained during the Network Challenge competition at SC'00 in November 2000. Our configuration for this competition consisted of eight Linux workstations on the SC'00 exhibition floor in Dallas, Texas, sending data across the wide area network to eight workstations (four Linux, four Solaris) at Lawrence Berkeley National Laboratory in California. Figure 4 illustrates the configuration. We used *striped* transfers during this competition, with a 2-gigabyte file partitioned across the eight workstations on the exhibition floor. Each workstation actually had four copies of its file partition. On each server machine, a new transfer of a copy of the file partition was initiated after 25% of the previous transfer was complete. Each new transfer creates a new TCP stream. At any time, there are up to four simultaneous TCP streams transferring data from each server in the cluster of eight workstations, for a total of up to 32 simultaneous TCP streams. Our current lack of data channel caching means that there are often fewer than four simultaneous streams transferring data on each host.

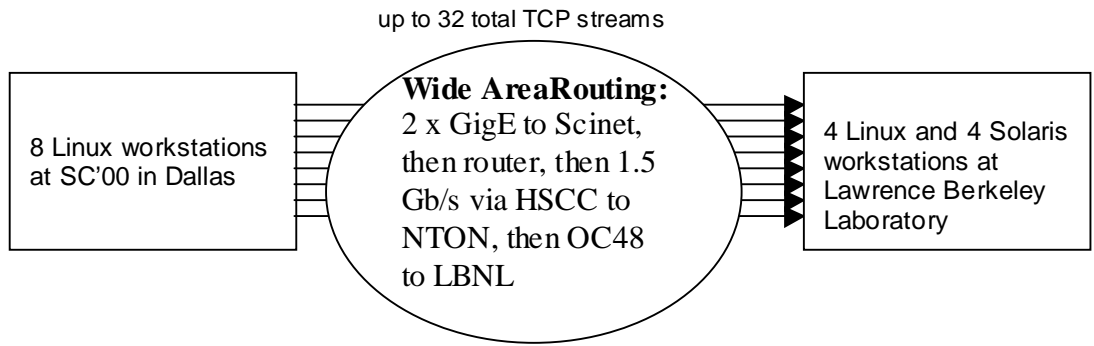

Figure 4: Experimental configuration for Network Challenge competition at SC'00.

Table 2 summarizes the results of our Network Challenge competition entry. We achieved a peak transfer rate of 1.55 gigabits/second over an interval of 0.1 seconds. This striped configuration was able to transfer data at a peak rate of 1.03 gigabits/second over an interval of 5 seconds. Over the hour-long period of our competition entry, we sustained an average data rate of 512.9 megabits per second. This corresponded to a total data transfer during that hour of 230.8 gigabytes, or a quarter of a terabyte.

Table 2: Network Challenge configuration and performance results

| | |
|---|---|
| Striped servers at source location | 8 |
| Striped servers at destination location | 8 |
| Maximum simultaneous TCP streams per server | 4 |
| Maximum simultaneous TCP streams overall | 32 |
| Peak transfer rate over 0.1 seconds | 1.55 Gbits/sec |
| Peak transfer rate over 5 seconds | 1.03 Gbits/sec |
| Sustained transfer rate over 1 hour | 512.9 Mbits/sec |
| Total data transferred in 1 hour | 230.8 Gbytes |

## 5 Replica Management

We next describe our second fundamental Data Grid service, that is, replica management. This component is responsible for managing the replication of complete and partial copies of *datasets,* defined as collections of files. Replica management services include:

- creating new copies of a complete or partial collection of files
- registering these new copies in a *Replica Catalog*
- allowing users and applications to query the catalog to find all existing copies of a particular file or collection of files
- selecting the ``best'' replica for access based on storage and network performance predictions provided by a Grid information service

The Globus replica management architecture is layered. At the lowest level we have a *Replica Catalog* that allows users to register files as logical collections and provides mappings between logical names for files and collections and the storage system locations of file replicas. Building on this basic component, we provide a low-level API that performs catalog manipulation and a higher-level Replica Management API that combines storage access operations with calls to low-level catalog manipulation functions. These APIs can be used by higher-level tools that select among replicas based on network or storage system performance, or that create (or delete) new replicas automatically at desirable locations.

**5.1 The Replica Catalog**

As mentioned above, the purpose of the replica catalog is to provide mappings between logical names for files or collections and one or more copies of those objects on physical storage systems. The catalog registers three types of *entries*: logical collections, locations, and logical files.

A *logical collection* is a user-defined group of files. We expect that users will often find it convenient and intuitive to register and manipulate groups of files as a collection, rather than requiring that every file be registered and manipulated individually. Aggregating files should reduce both the number of entries in the catalog and the number of catalog manipulation operations required to manage replicas.

*Location* entries in the replica catalog contain the information required for mapping a logical collection to a particular physical instance of that collection. The location entry may register information about the physical storage system, such as the hostname, port and protocol. In addition, it contains all information needed to construct a URL that can be used to access particular files in the collection on the corresponding storage system. Each location entry represents a complete or partial copy of a logical collection on a storage system. One location entry corresponds to exactly one physical storage system location. The location entry explicitly lists all files from the logical collection that are stored on the specified physical storage system.

Each logical collection may have an arbitrary number of associated location entries, each of which contains a (possibly overlapping) subset of the files in the collection. Using multiple location entries, users can easily register logical collections that span multiple physical storage systems.

Despite the benefits of registering and manipulating collections of files using logical collection and location objects, users and applications may also want to characterize individual files. For this purpose, the replica catalog includes optional entries that describe individual *logical files*. Logical files are entities with globally unique names that may have one or more physical instances. The catalog may optionally contain one logical file entry in the replica catalog for each logical file in a collection.

Figure 5 shows an example replica catalog for a climate modeling application. This catalog contains two logical collections with $CO_2$ measurements for 1998 and 1999. The

1998 collection has two physical locations, a partial collection on the hos jupiter.isi.edu and a complete collection on sprite.llnl.gov. The location entries contain attributes that list all files stored at a particular physical location. They also contain attributes that provide all information (protocol, hostname, port, path) required to map from logical names for files to URLs corresponding to file locations on the storage system. The example catalog also contains logical file entries for each file in the collection. These entries provide size information for individual files.

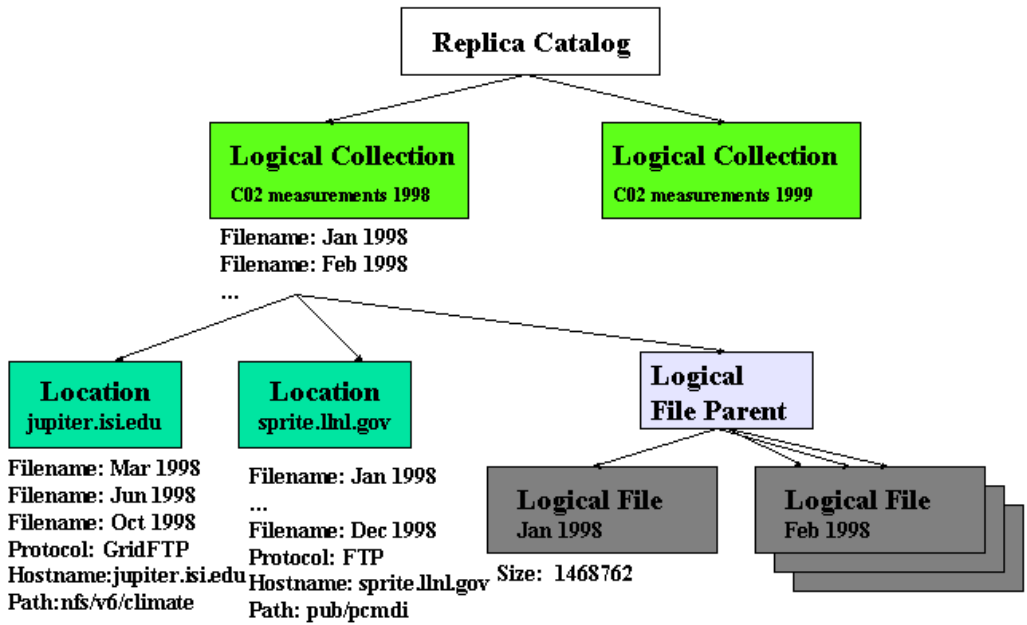

Figure 5: A Replica Catalog for a climate modeling application.

### 5.2 Replica Catalog API and Command Line Tool

We have implemented an API for low-level replica catalog manipulation as a C library called **globus_replica_catalog.c**. In addition, a straightforward command-line tool provides similar functionality. There are three types of operations on replica catalog entries. First, the API provides functions to create and delete catalog entries, for example, to register a new collection or location. Second, the API provides functions to add, list or delete individual attributes of a catalog entry. For example, as an experimental physics application produces new data files, the collection owner can register these files with the replica catalog by adding their names as attributes of existing logical collection and location entries. Third, the API provides functions to list or search catalog entries, including complex search operations that find all physical locations where a particular set of logical files is stored.

### 5.3 Replica Management API

The Replica Management API is a higher-level API that combines storage system operations with calls to low-level replica catalog API functions. Key concepts include the following:

- *Registration:* Registration operations add information about files on a physical storage system to existing location and logical collection entries. For example, when a long-running scientific experiment periodically produces new data files, these files are made available to users by registering them in existing location and collection entries.

- *Copying:* This operation *copies* a file between two storage systems that are registered as locations of the same logical collection and *updates* the destination's location entry to include the new file.

- *Publishing:* The publishing operation takes a file from a source storage system that is not represented in the replica catalog, *copies* the file to a destination storage system that is represented in the replica catalog, and *updates* the corresponding location and logical collection entries.

**5.4 Replica Management Architecture Implementation and Performance**

In this section, we present preliminary performance results for our prototype implementation of the Globus replica management architecture. We have implemented the low-level replica catalog manipulation API in C. The replica catalog itself is currently implemented as a Lightweight Directory Access Protocol (LDAP) directory, although future implementations may use relational databases. (The higher-level Replica Management API has not yet been implemented.) Our experimental replica catalog is a Netscape Directory Server version 4.12 LDAP directory configured with a cache size limit of 100 objects and a limit of 100 megabytes for caching database index files. This LDAP server runs on a 333 MHz Sun Sparc Ultra-5 workstation with 384 megabytes of memory running the SunOS version 5.7 operating system. The LDAP directory maintains an index on filename attributes of logical collection and location entries. A single client submits requests to the LDAP server in our tests.

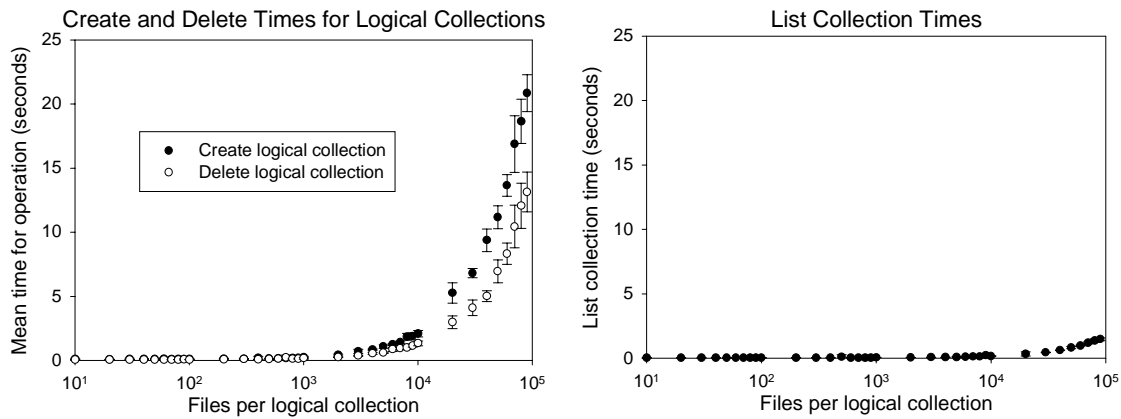

Figure 6: Microbenchmark performance results for low-level replica catalog API

Figure 6 shows microbenchmark performance results for three operations of the low-level replica catalog manipulation API: creating and deleting a logical collection and listing the contents of a collection entry. Each is graphed on the same scale as the size of collections increases from 1 file to 100,000 files. All measurements are run 10 times, with the graphs indicating mean times for the operations and error bars showing standard deviation. We include only graphs for operations on logical collections. (Graphs for operations on location entries show similar behavior.) The graph on the left indicates the time required to create and delete logical collection entries in the catalog, where each entry has one attribute per logical file name. As the number of files in a logical collection approaches 100,000, the create and delete times increases considerably, to approximately 20 seconds. This is due to the large number (tens of thousands) of filename attributes associated with these large logical collection entries. Although these times are relatively long, creation and deletion operations should be fairly rare. We also note that our experimental LDAP server runs on a relatively low-performance workstation.

We expect that list and search operations will be more common than create or delete operations. The graph on the right of Figure 6 shows the time to list the contents of a logical collection entry in the catalog. The list operation is much faster than create/delete operations, with times ranging from well under one second for small collections to approximately one second for collections with tens of thousands of files. Another operation expected to occur frequently is a search for all locations of a logical file. We are currently running experiments to measure search performance. Preliminary results for collections with one, two, four and eight locations show mean search times of approximately 1 second regardless of the number of locations per collection. Because we maintain an index on the filenames in the replica catalog, searches for all locations of a file appear to be efficient regardless of the number of locations.

**6 Conclusions**

We have argued that high-performance, distributed data-intensive applications require two fundamental services: *secure, reliable, efficient data transfer* and the ability to *register, locate, and manage multiple copies* of datasets. These two services can be used to build a range of higher-level capabilities, including reliable creation of a copy of a data collection at a new location, selection of the best replica for a data transfer operation based on performance, and automatic creation of new replicas in response to application demands.

We have presented our design and implementation of these two services. The GridFTP protocol implements extensions to FTP that provide GSI security and parallel, striped, partial, and third-party transfers, while the Globus replica management architecture supports the management of complete and partial copies of datasets. Performance studies of both components provide promising results.

These and other Globus Toolkit services are being applied by ourselves and others in a variety of large-scale application projects, including the Particle Physics Data Grid (www.ppdg.net), Earth Systems Grid, Grid Physics Network (www.griphyn.org), and European Data Grid (grid.web.cern.ch/grid) projects. Experience with these applications will motivate further refinements and additions to the services described here. We are already planning extensions, such as automated replica management, community-based access control, automated buffer size negotiation, and server-side data reduction [14].

**Acknowledgements**

We are grateful to Marcus Thiebaux and Soonwook Hwang for their work characterizing the performance of LDAP servers; to Brian Toonen, who helped to optimize the GridFTP code; to Gail Pieper, Laura Pearlman and Ewa Deelman for comments on this paper; and to the many colleagues in the sciences who helped us understand their requirements. This work was supported in part by the Mathematical, Information, and Computational Sciences Division subprogram of the Office of Advanced Scientific Computing Research, U.S. Department of Energy, under Contract W-31-109-Eng-38, and by the National Science Foundation.